\documentclass[conference,compsoc]{IEEEtran}
\ifCLASSOPTIONcompsoc
  \usepackage[nocompress]{cite}
\else
  \usepackage{cite}
\fi
\ifCLASSINFOpdf
\else
\fi
\usepackage{url}
\urldef{\mailsa}\path|{cise,leonhard,andra,ozgu,griff}@simula.no|
\urldef{\mailsb}\path|leo@lww.at|

\usepackage{soul}
\usepackage{amssymb}
\usepackage{graphicx}
\usepackage{subfig}
\usepackage{multirow}
\usepackage{hhline}
\usepackage[printonlyused]{acronym}
\usepackage[table,x11names]{xcolor}
\usepackage{color, colortbl}
\newcolumntype{g}{>{\centering\columncolor{lightgray}}p{1cm}}

\usepackage{tabularx}

\usepackage{array}
\newcolumntype{x}[1]{>{\centering\arraybackslash\hspace{0pt}}p{#1}}

\usepackage{siunitx}
\DeclareSIUnit\bps{\bit\per\second}
\DeclareSIUnit\kbps{\kilo\bit\per\second}
\DeclareSIUnit\Mbps{\mega\bit\per\second}
\DeclareSIUnit\Bps{\byte\per\second}
\DeclareSIUnit\kBps{\kilo\byte\per\second}
\DeclareSIUnit\MBps{\mega\byte\per\second}
\DeclareSIUnit\kiBps{\kibi\byte\per\second}
\DeclareSIUnit\MiBps{\mebi\byte\per\second}

\acrodef{3G}{Third Generation}
\acrodef{4G}{Fourth Generation}
\acrodef{ADB}{Android Debug Bridge}
\acrodef{ASCI}{Advertising Standards Council of India}
\acrodef{ASN}{Autonomous System Number}
\acrodef{CDN}{Content Delivery Network}
\acrodef{DL}{Downlink}
\acrodef{DNS}{Domain Name Service}
\acrodef{EaaS}{Experiment as a Service}
\acrodef{ECDF}{Empirical Cumulative Distribution Function}
\acrodef{HTTP}{Hyper Text Transfer Protocol}
\acrodef{ICMP}{Internet Control Message Protocol}
\acrodef{ISP}{Internet Service Provider}
\acrodef{IXP}{Internet Exchange Point}
\acrodef{LTE}{Long Term Evolution}
\acrodef{MBB}{Mobile Broadband}
\acrodef{MONROE}{Measuring Mobile Broadband Networks in Europe}
\acrodef{MNO}{Mobile Network Operator}
\acrodef{MSC}{Message Sequence Chart}
\acrodef{NDT}{Network Diagnostic Tool}
\acrodef{QoE}{Quality of Experience}
\acrodef{QoS}{Quality of Service}
\acrodef{RMBT}{RTR Multithreaded Broadband Test}
\acrodef{RTT}{Round-Trip Time}
\acrodef{SIM}{Subscriber Identity Module}
\acrodef{SLA}{Service-Level Agreement}
\acrodef{TCP}{Transmission Control Protocol}
\acrodef{UDP}{User Datagram Protocol}
\acrodef{UE}{User Equipment}
\acrodef{UL}{Uplink}
\acrodef{NRA}{National Regulatory Authority}

\begin{document}

\bstctlcite{IEEEexample:BSTcontrol}

%
\title{MONROE-Nettest: A Configurable Tool for Dissecting Speed Measurements in Mobile Broadband Networks}

\author{\IEEEauthorblockN{Cise Midoglu, 
Leonhard Wimmer, 
Andra Lutu, 
\"Ozg\"u Alay 
and Carsten Griwodz
}
\IEEEauthorblockA{
Simula Research Laboratory\\
\{cise,leonhard,andra,ozgu,griff\}@simula.no
}
}

\maketitle


\begin{abstract}\label{sec:abstract}
As the demand for mobile connectivity continues to grow, there is a strong need to evaluate the performance of Mobile Broadband (MBB) networks.
In the last years, mobile ``speed", quantified most commonly by data rate, gained popularity as the widely accepted metric to describe their performance.
However, there is a lack of consensus on how mobile speed should be measured.
In this paper, we design and implement MONROE-Nettest to dissect mobile speed measurements, and investigate the effect of different factors on speed measurements in the complex mobile ecosystem.
MONROE-Nettest is built as an Experiment as a Service (EaaS) on top of the MONROE platform, an open dedicated platform for experimentation in operational MBB networks.
Using MONROE-Nettest, we conduct a large scale measurement campaign and quantify the effects of measurement duration, number of TCP flows, and server location on measured downlink data rate in 6 operational MBB networks in Europe. Our results indicate that differences in parameter configuration can significantly affect the measurement results. 
We provide the complete MONROE-Nettest toolset as open source and our measurements as open data.
\end{abstract}



%
\IEEEpeerreviewmaketitle

\section{Introduction}\label{sec:introduction}
The use of mobile networks has exploded over the last few years due to the immense popularity of powerful mobile devices combined with the availability of high-capacity 
3G/4G mobile networks.
Forecasts indicate that global mobile data traffic will increase sevenfold between 2016 and 2021, reaching $48$ EB per month by 2021 and summing up to a total of $17\%$ of all IP traffic~\cite{Cisco2017}.
This reflects the insatiability of mobile data consumers and the rapid boost in demand for faster mobile connectivity.
Given the increasing importance of \ac{MBB} networks and the expected growth in mobile traffic, there is a strong need for better understanding the fundamental characteristics of \ac{MBB} networks and the services that they can provide.

Ensuring a smooth mobile experience for customers is key to business success 
for \acp{MNO} and mobile ``speed" is a foolproof marketing resource (i.e., it is straightforward for any customer to understand that, when it comes to mobile connectivity, higher speeds are better).
Consequently, in the recent years, the term mobile speed has become the major indicator of the \ac{MNO}'s performance.
For instance, the broadband testing company Ookla gives out \textit{Fastest Broadband and Mobile Network Awards} every year, where they use a ``Speed Score" calculated using \ac{DL} data rate and \ac{UL} data rate~\cite{speedtestawards}. Similarly, OpenSignal
compiles yearly reports on the state of mobile networks around the world with award tables, where average \ac{DL} data rate is used as the ``speed" indicator~\cite{web_OpenSignal_reports}.
The rankings and reports from such performance monitoring entities steer the public opinion in the \ac{MBB} market and impact customer behavior, since the simple concept of speed resonates well with end-users.

Despite the apparent simplicity of the term, there is a lack of consensus about how to measure mobile speed (e.g. \ac{DL} and \ac{UL} data rates) accurately and consistently.
For example, all of the commercial mobile speed measurement tools (e.g. Speedtest and OpenSignal) are proprietary and the measurement methodologies are not open (i.e. only high-level explanations in the tool descriptions and FAQ pages are provided). 
This lack of transparency can result in controversy, as in the case of Ookla designating \textit{Airtel} as India's fastest 4G network in 2016, which was a move questioned by another operator \textit{Reliance Jio} and subsequently caused a complaint made to the \ac{ASCI}~\cite{web_Ookla_India}.
To eliminate this controversy as well as verify that the services paid for by the end-users are actually provided by their \acp{MNO}, \acp{NRA} have come up with their own crowdsourced solutions. 
Notable examples of regulatory tools include the FCC Speed Test~\cite{web_FCCSpeedTest} in the US, RTR-Nettest in Austria~\cite{web_RTRNetztest, Midoglu2016}, and similar tools that are based on the \ac{RMBT} in other countries such as Croatia, Czech Republic, Slovakia, Slovenia, and Norway. 
Despite their fundamental similarities, different tools use different parameter sets during measurement (e.g. measurement duration, number of \ac{TCP} flows, and server location)~\cite{Goel2016}, leading to significant differences in their reported results\cite{khatouni2017itc}.
Therefore, it is important to understand how different parameters 
influence mobile speed measurements.

In this paper, we design and implement an open source, configurable measurement tool, MONROE-Nettest, that integrates the measurement functionality of existing tools to understand how different parameters 
influence mobile speed measurements. 
MONROE-Nettest is built as an \ac{EaaS} on top of the \ac{MONROE}\footnote{\url{https://www.monroe-project.eu/}} platform, Europe's first and only open dedicated platform for experimentation in operational \ac{MBB} networks. 
\ac{MONROE} makes it possible to conduct a wide range of repeatable measurements in the same location, using the same devices/modems, for multiple operators at the same time, and from an end-user perspective.
Therefore, MONROE-Nettest enables the empirical analysis of mobile speed measurements through large-scale experimentation in operational \ac{MBB} networks. 
We demonstrate a use case of MONROE-Nettest by quantifying the effects of several parameters on \ac{DL} data rate through a systematic large-scale campaign over 6 operational mobile networks in Europe.
Our results show that 
differences in configuration can significantly affect measurement results, and that datasets from different tools need to be baselined in a controlled framework for comparability.

\section{Background and Related Work}\label{sec:relatedwork}

\ac{UDP}-based measurements are widely used to explore the available capacity for a given network~\cite{Ferlin2014}. 
In our previous work, we presented such an end-to-end active measurement method \cite{Bideh2015}. 
However, \ac{TCP} is more representative of the \textit{share} of the bottleneck capacity that the end users experience, especially with short-lived application traffic which is most common in today's Internet. 
Therefore, speed measurement tools adopt such a user oriented approach for measuring data rate and they use \ac{TCP}-based testing with multiple parallel flows. 

Speed measurements have been well studied in broadband networks~\cite{Bauer2010, Canadi2012} where a comparison of different methodologies is provided for browser-based testing. However, many concepts that are already known in fixed broadband networks are harder to track, evaluate, and more importantly, quantify in \ac{MBB} networks~\cite{khatouni2017itc}. 
Therefore, it is important to revisit speed measurements considering the complexity of \ac{MBB} networks.
When it comes to mobile speed measurements, crowdsourced tools such as \ac{NDT}~\cite{web_NDT}, MobiPerf~\cite{Nikravesh2014} and Netalyzr~\cite{Kreiblich2010, Kreiblich2011} are commonly used to collect measurements from a large number of \ac{MBB} users. A recent survey provides a comprehensive summary of crowdsourced tools that are intended for end-user measurements of mobile networks through smartphone applications~\cite{Goel2016}. However, despite its breadth and relative depth, it does not provide an empirical comparative analysis of the different measurement methodologies.

While crowdsourced approaches provide a large number of vantage points spread throughout different regions and access to numerous networks and user equipment, repeatability is challenging to achieve and one can only collect measurement data at users' own will.  Dedicated infrastructures, on the other hand, can provide active, systematic measurements that can be run at regular intervals over long time periods.
Moreover, they allow fair assessment of each network by following the same measurement methodology and running the measurements under similar conditions (e.g., same device, same operating system, same environment). There are many dedicated testbeds for measuring broadband networks, such as PlanetLab~\cite{Chun2003}, GENI~\cite{Berman2014} 
and Fed4FIRE~\cite{web_fed4fire},
However, support for \ac{MBB} is limited to Fed4FIRE, where none of its associated testbeds such as NITOS~\cite{nitos} and WiLAB~\cite{wilab}, support experimentation in \textit{operational} MBB networks. 
To this end, \ac{MONROE}~\cite{alay2016wowmom, alay2017mobicom, Peon2017} is the only dedicated testbed that enables large-scale end-to-end measurements in operational \ac{MBB} networks. 
In this paper, building on the \ac{MONROE} platform, we design and implement a configurable tool that integrates the measurement functionalities of existing crowdsourced applications. MONROE-Nettest enables the empirical analysis of mobile speed measurements through large-scale experimentation in operational \ac{MBB} networks. The tool is provided as \ac{EaaS} and allows for  
any interested party to investigate different aspects of speed measurements using \ac{MONROE}. 
\section{Tool Design and Implementation}
\label{sec:tool}

In this section, we first provide an overview of the \ac{MONROE} platform, and then detail the design and implementation of MONROE-Nettest. 

\subsection{\ac{MONROE} Platform}\label{sec:MONROE}
\ac{MONROE}\cite{alay2017mobicom} is a European transnational open platform, and
the first open access hardware-based platform for independent, multi-homed, and
large-scale \ac{MBB} measurements on commercial networks. 
The platform comprises a set of 150 nodes, both mobile (e.g., operating in delivery trucks and on board public transport vehicles, such as trains or busses) 
and stationary (e.g., volunteers hosting nodes in their homes). 
Nodes are multi-homed
to 3 different \ac{MBB} operators using commercial-grade subscriptions in several countries in Europe (Italy, Norway, Spain, Sweden, Portugal, Greece and UK).

Each \ac{MONROE} node integrates two small programmable computers (PC Engines APU2 board interfacing with three 3G/4G MC7455 miniPCI express modems using LTE CAT6 and one WiFi modem). 
The software on the node is based on Debian GNU/Linux ``stretch" distribution and each node collects metadata from the modems, such as carrier, technology, signal strength, GPS location and sensor data.
This information is made available to the experimenters during 
execution. 
Experiments running on the platform uses Docker containers (light-weight virtualized environment) to provide agile reconfiguration. 

User access to the platform resources is through a web portal, which allows an authenticated user to use the \ac{MONROE} scheduler to deploy experiments. This enables exclusive access to nodes (i.e., no two experiments run on the node at the same time).
The results from each experiment are periodically transferred from the nodes to a repository at a back-end server, while the \ac{MONROE} scheduler also sets data quotas to ensure fairness among users.

\ac{MONROE} allows us to eliminate the noise from measurements, by providing a controlled environment to conduct repeatable and reproducible experiments.
In contrast to crowdsourced tools which cannot produce datasets for performance characterization of \acp{MNO} due to the amount of noise in their results or because of app permission requirements~\cite{Kreiblich2010}, \ac{MONROE} provides a clean dataset collected from identical devices that require no maintenance on the part of the end user.

All software components used in the platform are open source and available online.\footnote{\url{https://github.com/MONROE-PROJECT/}\label{foot:MONROEgit}}

\subsection{MONROE-Nettest Client}

\subsubsection*{Client Core}\label{sec:rmbtcore}

We choose \ac{RMBT}\cite{web_RTRdoc} as the codebase for our implementation since it is used by a number of \acp{NRA} in Europe for their crowdsourced measurement applications (see Section \ref{sec:introduction}). However, the current \ac{RMBT} core is in Java, as this is the primary programming language for Android smartphones. 
Our goal is to provide a light-weight implementation which can also be run on devices with relatively low resources. Using a lower-level language such as C enables this, while at the same time allowing direct access to socket functions provided by the standard library. 
C also enables cross-platform compatibility, which allows our tool to be adaptable to other testbeds with minimal effort. Our client core is open-source, and can also be used as a native library in different tools or mobile applications.\footnote{https://github.com/lwimmer/rmbt-client} 

\begin{figure*}
\centering
\subfloat[Pre-test \ac{DL}]{
\includegraphics[width=0.119\textwidth]{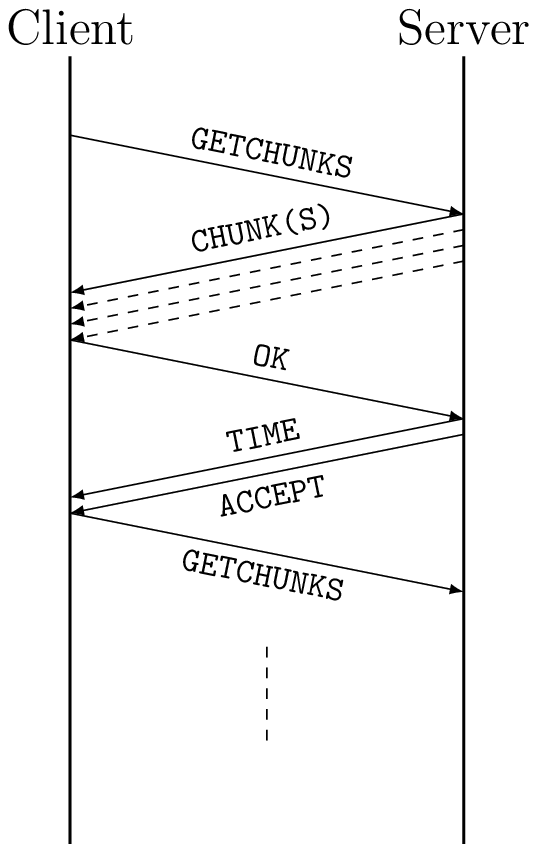}}
\subfloat[Ping]{
\includegraphics[width=0.119\textwidth]{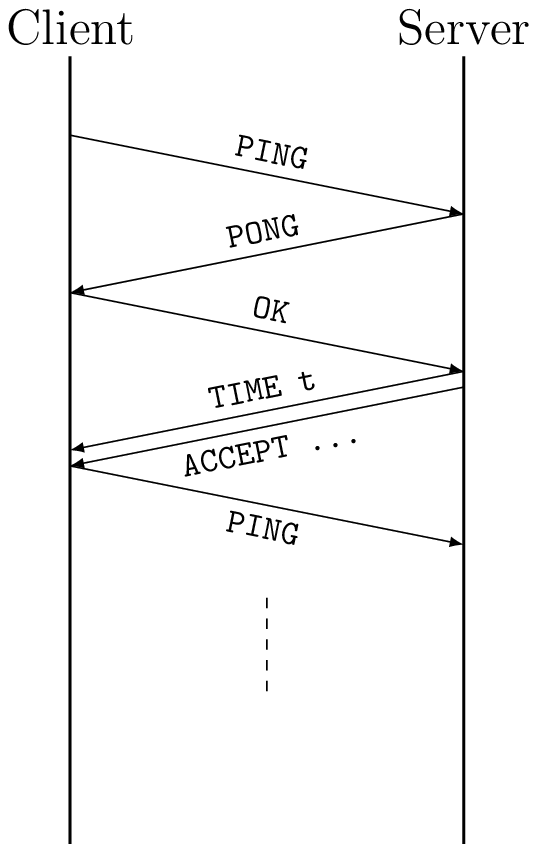}}
\subfloat[\ac{DL}]{
\includegraphics[width=0.119\textwidth]{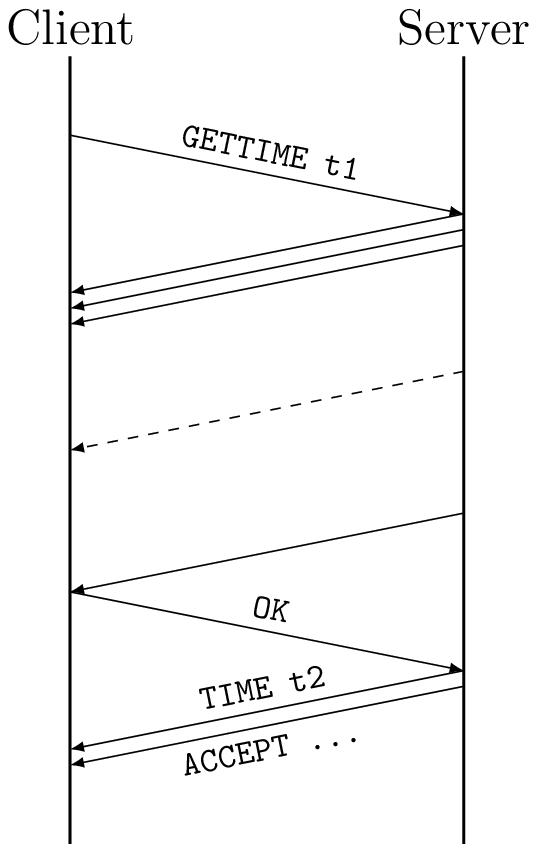}}
\subfloat[Pre-test \ac{UL}]{
\includegraphics[width=0.119\textwidth]{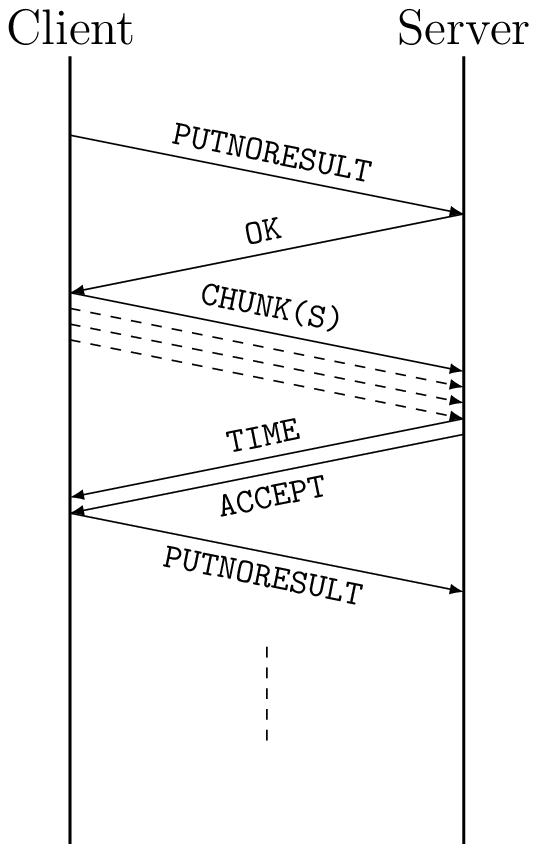}}
\subfloat[\ac{UL}]{
\includegraphics[width=0.119\textwidth]{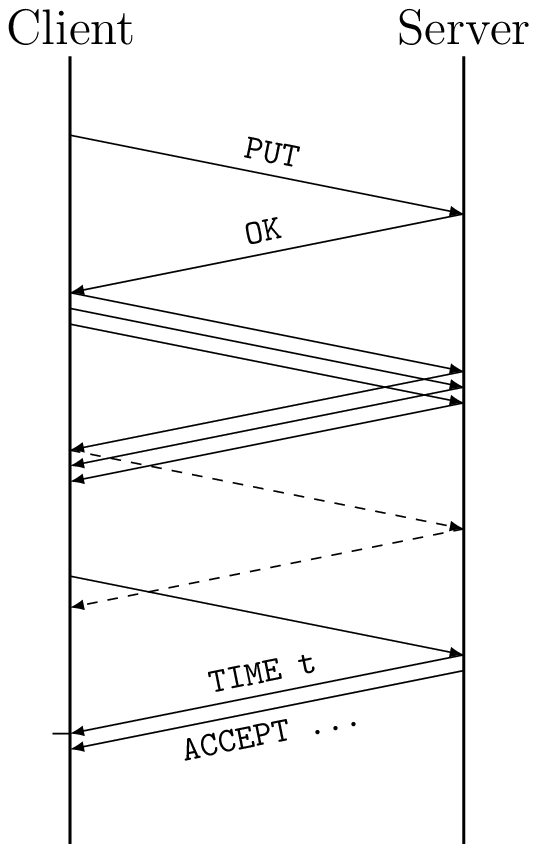}}
\caption{\acsp{MSC} for the different Nettest phases.}
\label{fig:timing-nettest}
\end{figure*}

We follow the \ac{RMBT} measurement flow. 
Figure~\ref{fig:timing-nettest} illustrates the \acp{MSC} for different measurement phases.
In the pre-test \ac{DL} phase, for each flow, the client requests data in the form of chunks that double in size for each iteration.
In the ping phase, the client sends $n$ small \ac{TCP} packets (ASCII \texttt{PING}), to which the server replies with an ASCII \texttt{PONG}.
The number of pings are configurable.
In the \ac{DL} phase, for each flow, the client continuously requests and the server continuously sends data streams consisting of fixed-size packets (chunks).
Pre-test \ac{UL} phase is similar to the pre-test \ac{DL} phase, except that the client sends chunks that double in size for each iteration.
In the \ac{UL} phase, for each flow, the client continuously sends data and the server continuously receives. For each chunk received, the server sends a timestamp indicating when it received the chunk.

In addition to the original \ac{RMBT} implementation, MONROE-Nettest samples high resolution time series of data rate (per chunk). It also collects \ac{TCP}-related information for each socket, such as retransmissions, \ac{RTT}, slow-start threshold, and window sizes using the \texttt{TCP\_INFO} socket option from the linux kernel (the granularity can be specified as input).

Configurable parameters of the client include the number of flows for \ac{DL} and \ac{UL}, measurement durations for \ac{DL}, \ac{UL} and the pre-tests, and the measurement server. The full list of configuration parameters and default values can be found in the experiment repository.\footnote{\url{https://github.com/MONROE-PROJECT/Experiments/tree/master/experiments/nettest}\label{foot:container}}

For calculating the data rate, the client uses an aggregation of all flows, with a granularity of one data chunk (which is also a configurable parameter).
Let $n$ be the number of \ac{TCP} flows used for the measurement and $F := \{1, \dots, n\}$ be the set of these flows.
All transmissions start at the same time, which is denoted as relative time $0$.
For each \ac{TCP} flow $f\!\in\!F$, the client records the relative time $t_f^{(i)}$
and the total amount $b_f^{(i)}$ of data received in Bytes on this flow (per chunk), from time 0 to $t_f^{(i)}$
for successive values of $i$, starting with $i := 1$ for the first chunk received.
For each \ac{TCP} flow$\:f\!\in\!F\!$, the time series begins with $t_f^{(0)}:=0$ and $b_f^{(0)}:=0$, where $m_f$ is the number of pairs $\Big(t_f^{(i)},b_f^{(i)}\Big)$ which have been recorded for flow $f$.
\begin{equation}
t^* := \min\big(\{t_f^{(m_f)} |\: \forall\:f\!\in\!F \}\big)
\label{eq:t_star}
\end{equation}
\begin{equation}
\forall\:f\!\in\!F\!: i_f := \min\big(\{i\!\in\!\mathbb{N} \:|\: 1 \leqslant i \leqslant m_f \land t_f^{(i)} \geqslant t^* \}\big)
\end{equation}
$i_f$ being the index of the chunk received on flow $f$ at $t^*$ or right after $t^*$.
Then the amount $b_f$ of data received over \ac{TCP} flow $f$ from time 0 to time $t^*$ is approximately
\begin{equation}
b_f :\approx b_f^{(i_f-1)} + \frac{t^* - t_f^{(i_f-1)}}{t_f^{(i_f)} - t_f^{(i_f-1)}} (b_f^{(i_f)}-b_f^{(i_f-1)})
\label{eq:b_f}
\end{equation}
The data rate for all \ac{TCP} flows combined is given by Eq.\ref{eq:R}, where $R$ is used as the final reported data rate.
\begin{equation}
R := \frac{1}{t^*} \sum_{f=1}^n b_f
\label{eq:R}
\end{equation}

One remark here is that we can get both the application data rate (e.g. including SSL overhead if enabled), and the transport capacity using the MONROE-Nettest. 
The former is provided by the Nettest core using Equation \ref{eq:R} for each measurement, where the latter can be calculated using the detailed \texttt{TCP\_INFO} output. 
Another important remark in connection with the slow-start phase of \ac{TCP} is that, in our current implementation, we are including this phase in the calculation. However, it is known that some of the existing tools 
cut out the \ac{TCP} slow start, which yields a more optimistic data rate estimate.

\subsubsection*{Experiment Container}
MONROE-Nettest is a wrapper around the client core explained in the previous segment, making it available for testbed experimentation. 
It runs as a Docker container\textsuperscript{\ref{foot:container}}, and combines a number of functionalities. 
First, it establishes that metadata information is available (if the experiment is run on \ac{MONROE} nodes), then it runs a traceroute against the selected measurement server, after which it runs the client core. 

The MONROE-Nettest container produces $4$ output files:
\textbf{\textit{summary.json}} presents the result fields such as calculated \ac{DL} and \ac{UL} data rate, and the median \ac{TCP} payload RTT. It also includes configuration-related fields such as the test identifier, basic input parameters, main timestamps, status, server and connection information, and metadata.
\textbf{\textit{flows.json}} includes the raw time series for each \ac{TCP} flow, which are sustained through all stages (initialization, pre-test \ac{DL}, ping, \ac{DL}, pre-test \ac{UL}, \ac{UL}). 
\textbf{\textit{stats.json}} includes samples of the socket option \texttt{TCP\_INFO} for each \ac{TCP} flow, in the given granularity. 
\textbf{\textit{traceroute.json}} includes the results from the traceroute towards the selected measurement server. If the \ac{ASN} per-hop is not available from the traceroute, this information is added via a lookup.

The advantage of running MONROE-Nettest over running the client core directly is the possibility to use the \textit{multi-config} option, which enables sets of experiments with different configuration parameters to be executed at once.

The MONROE-Nettest container can be run within or outside the scope of \ac{MONROE}.

\subsubsection*{Scheduler Template}
As mentioned in Section \ref{sec:MONROE}, \ac{MONROE} provides a scheduler with a web interface for external experimenters.\footnote{\url{https://www.monroe-system.eu/}, instructions on how to use the scheduler for certified users can be found in the user manual.} A template is prepared to run MONROE-Nettest from this web interface with a single click, using the default parameters. 
The configuration parameters can be modified at will.

\subsection{MONROE-Nettest Server}

In order to keep compatibility with the \ac{RMBT}, we use the server code from the open-source Open-RMBT project\cite{web_RMBTserver}. Our only change is to disable the token check so that the measurement server can be used without previous token/secret exchange. Therefore, the MONROE-Nettest container can be run against any existing \ac{RMBT} server whose token check is disabled.

The \ac{MONROE} testbed readily provides \num{4} servers in Germany, Norway, Spain, and Sweden. Further, our source code is open so that any institution can host a MONROE-Nettest server at their premises.\footnote{Instructions on how to set up a MONROE-Nettest server can be found under\textsuperscript{\ref{foot:container}}.}
\section{Measurement Results and Evaluation}\label{sec:results}

In this section, we focus on the \ac{DL} data rate feature of MONROE-Nettest and demonstrate the capabilities of our tool by quantifying the impact of different parameters on \ac{DL} data rate for operational MBB networks.

\begin{table*}
\centering
\caption{MONROE-Nettest measurement campaigns.}
\begin{tabular}{|c|c|c|c|c|c|c|}
\hline
\textbf{ID}&
\textbf{Batches}&\textbf{Operator}&\textbf{Clients}&\textbf{Servers}&\textbf{Flows}&\textbf{Duration}\\
\hline
1 & \num{1215} & 3 NO, 3 SE & 10 NO, 10 SE & 1 NO, 1 SE & 1,3,5,7 & \SI{15}{\second} \\
2 & \num{910} & 3 NO, 3 SE & 2 NO, 2 SE & 1 NO, 1 SE & 1,3,4,5,7,9 & \SI{15}{\second} \\
3 & \num{87} & 2 NO & 1 NO & 1 SE, 1 DE & 3,5,7,9 & \SI{15}{\second} \\
\hline
\end{tabular}
\label{tab:measurements}
\end{table*}

\subsection{Measurement Setup}
\label{sec:measurementsetup}
For our measurement campaign, we leverage stationary \ac{MONROE} nodes and opt out 
of using mobile \ac{MONROE} nodes, in order to eliminate any mobility related impact on our results.
Considering the fact that mobile subscriptions have limited data quotas and running active measurements under multiple configurations for a relatively long duration of time consumes large data volumes, we focus on Scandinavian countries, where the SIM quotas are relatively higher.
More specifically, we use \num{10} nodes distributed in Oslo, Norway and \num{10} nodes distributed in Karlstad, Sweden.
We measure a total of \num{6} different commercial networks in 4G.
We run measurements against servers in Norway (Oslo), Sweden (Karlstad), and Germany (Falkenstein).
Table~\ref{tab:measurements} lists our measurement campaigns.
Campaign \num{1} focuses on a smaller number of flows and \num{2} servers, with extended vantage points to eliminate any location specific artifacts on the results, whereas Campaigns \num{2} and \num{3} jointly provide a deeper  
analysis of the number of flows and server location, with less vantage points, all using \SI{15}{\second} measurement duration.
Overall, we run more than \num[group-minimum-digits = 4]{2000} batches, corresponding to \num[group-minimum-digits = 4]{10000} individual experiment runs. 

\subsection{Number of Flows and Measurement Duration}

\begin{figure*}
\centering
\subfloat[Campaign 1, overall.]{
\includegraphics[width=0.33\textwidth]{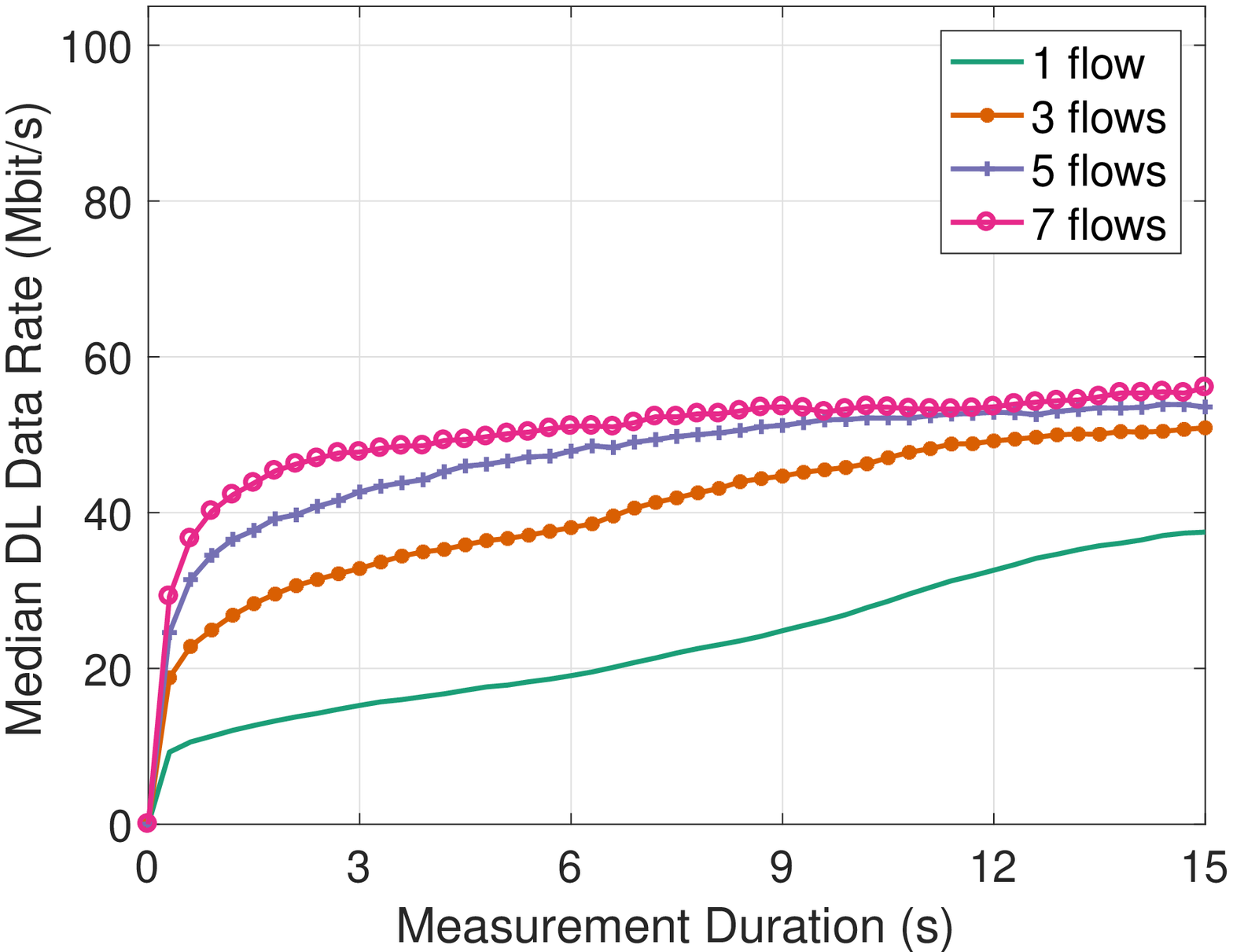}
\label{fig:numflowsduration_a}}
\subfloat[Campaign 1, op1-SE.]{
\includegraphics[width=0.33\textwidth]{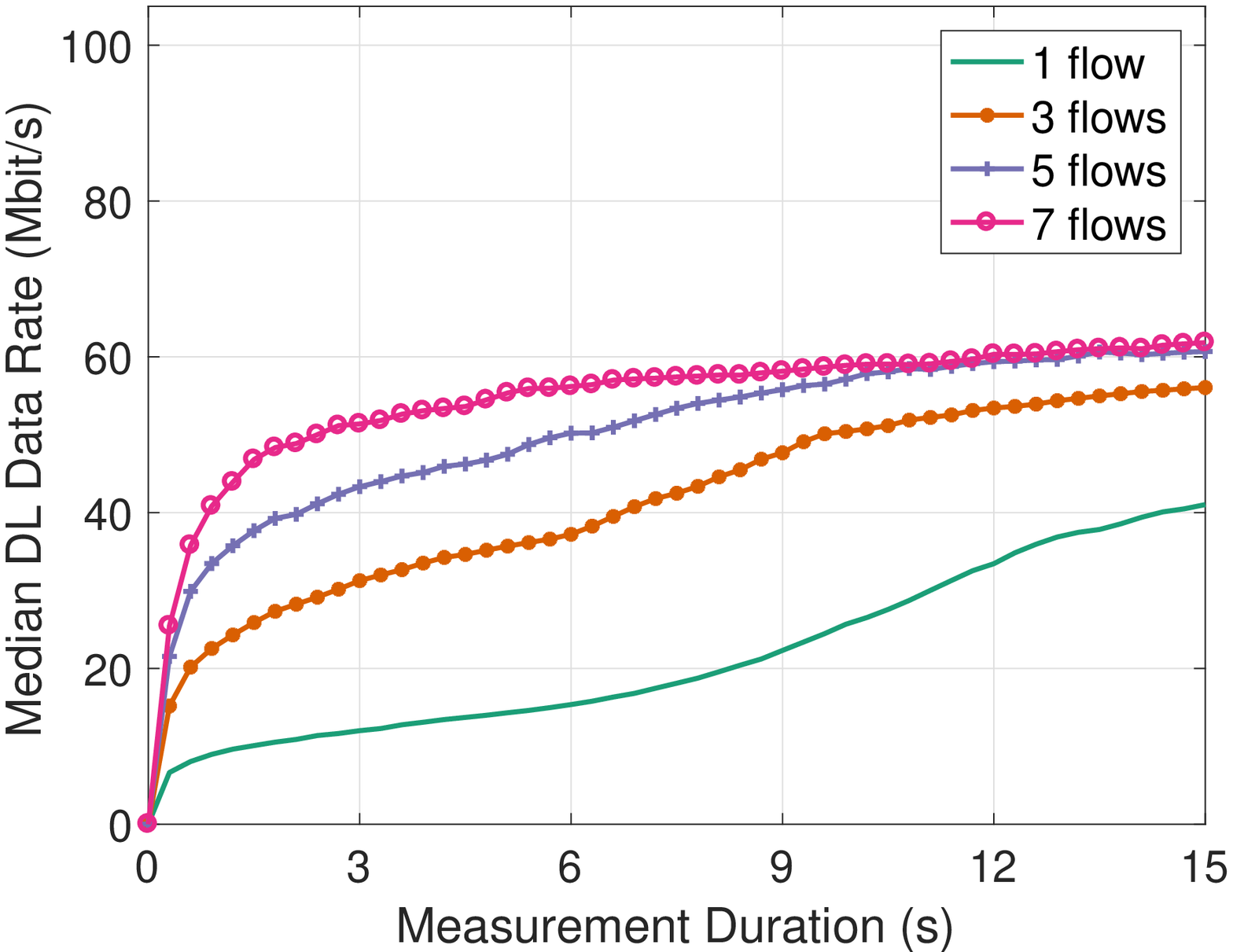}
\label{fig:numflowsduration_b}}
\subfloat[Campaign 1, op2-SE.]{
\includegraphics[width=0.33\textwidth]{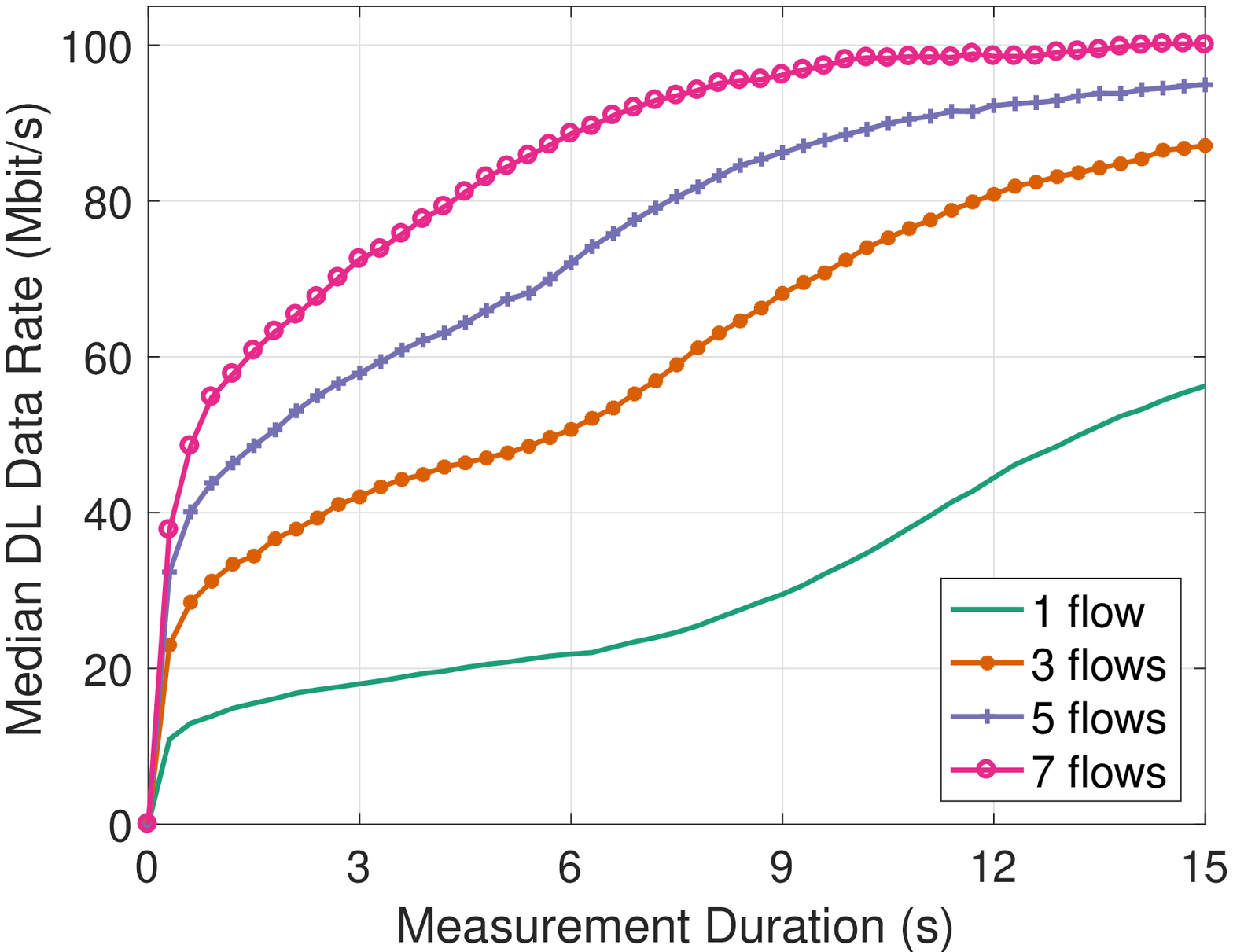}
\label{fig:numflowsduration_c}}
\caption{Effect of number of flows and measurement duration.}
\label{fig:numflowsduration}
\end{figure*}

First, we investigate the effect of number of flows and measurement duration on reported \ac{DL} data rate.
Figure~\ref{fig:numflowsduration} shows reported median \ac{DL} data rate vs. measurement duration for different number of flows, from Campaign \num{1}.
For each run, we first compute the reported data rate using Equation \ref{eq:t_star} for every \SI{10}{\milli\second} between \SI{0}{\second}-\SI{15}{\second} as $t^*$. We then used the median of these values, per $t^*$, to plot the curves in the figure.
We observe a general trend of increasing reported data rate with increased number of flows (Figure~\ref{fig:numflowsduration_a}). The trend is also persistent in the smaller scale, when we only consider nodes from a single operator against a measurement server in the same country (Figure~\ref{fig:numflowsduration_b}-\ref{fig:numflowsduration_c}).
We further observe that the maximum supported \ac{DL} data rate of a network (as an artifact of operator coverage and provisioning) impacts the spread between different number of flows.

Next, we try to \textit{quantify} these effects using Campaign \num{2}, which includes $6$ different number of flows: \num{1}, \num{3}, \num{4}, \num{5}, \num{7}, \num{9}. 
We evaluate the similarity of time series using Euclidean distance, which has been shown to be applicable in this context~\cite{Serra2014}.
Table~\ref{tab:numflowsduration} presents the percentage of the median data rate at \SI{15}{\second} (assumed to be the saturation data rate) that the curves capture at \SI{2}{\second}, \SI{4}{\second}, \SI{6}{\second}, \SI{8}{\second}, and \SI{10}{\second}. We observe that while with \num{9} flows, it takes \num{2} seconds to reach more than 92\% of the saturation data rate, it takes \num{6} seconds to reach the percentage with \num{3} flows. Note that there is a trade-off between the accuracy in estimation, which increases with the number of flows, and data consumption, which increases with the time spent by each flow at a ``saturated" state. Therefore, to avoid consuming unnecessary data quota while accurately capturing the data rate, there is a need for optimizing the number of flows and test duration. The optimal value of these parameters, however, might vary according to operator, technology, and coverage. 

\begin{table}[h!]
\centering
\caption{Quantifying the effect of number of flows and measurement duration.}
\begin{tabular}{|c|c|c|c|c|c|}
\hline
&\multicolumn{5}{|c|}{\textbf{\% of Saturation Data Rate}}\\
\hline
\textbf{Flows}&\textbf{2s}&\textbf{4s}&\textbf{6s}&\textbf{8s}&\textbf{10s}\\
\hline
1 & $67.5$ & $78.6$ & $82.5$ & $84.9$ & $89.4$ \\
3 & $73.3$ & $88.6$ & $92.9$ & $96.2$ & $98.2$\\
4 & $78.3$ & $91.6$ & $95.6$ & $97.5$ & $98.2$ \\
5 & $83.9$ & $94.2$ & $96.9$ & $97.6$ & $99.0$\\
7 & $91.0$ & $96.4$ & $98.3$ & $99.0$ & $99.1$ \\
9 & $92.8$ & $95.6$ & $97.8$ & $98.1$ & $99.0$ \\
\hline
\end{tabular}
\label{tab:numflowsduration}
\end{table}

\begin{figure}
\centering
\includegraphics[width=0.8\columnwidth]{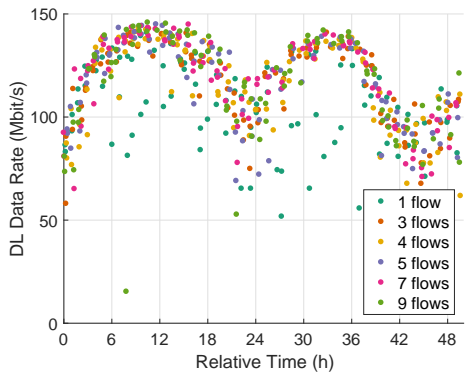}
\caption{Daily patterns in reported data rate.}
\label{fig:operatorextra}
\end{figure}

We further observe daily data rate patterns for some operators, indicating that this spread might not be constant throughout a day.
Figure~\ref{fig:operatorextra} shows the scatter plot of data rate vs. relative time for a Norwegian operator (op4), for the first \num{50} hours of Campaign \num{2}.
There is a clear \SI{24}{\hour} trend for all values of the number of flows above \num{1}.
This confirms that the optimal value for the number of flows and test duration needs to take also the daily patterns into account.

\subsection{Server Location} Next, we investigate the effect of server location by comparing measurements against servers in different countries, using Campaigns \num{2} and \num{3}. Table~\ref{tab:serverlocation} presents the Euclidean distance between time series of reported median \ac{DL} data rate measured by clients in Norway using a Norwegian SIM (op4) against servers in different countries (Germany vs. Sweden, and Norway vs. Sweden), for different number of flows (\num{3},\num{5},\num{7},\num{9}). The values are normalized according to the median value from the Sweden curve (results against the server in Sweden).

\begin{table}[h!]
\centering
\caption{Euclidean distance between time series.}
\begin{tabular}{|c|c|c|c|c|}
\hline
\textbf{Flows}&\textbf{Germany - Sweden}&\textbf{Norway - Sweden}\\
\hline
3 & $73.4\%$ & $46.5\%$\\
5 & $61.2\%$ & $37.3\%$\\
7 & $54.8\%$ & $32.8\%$\\
9 & $40.0\%$ & $26.7\%$\\
\hline
\end{tabular}
\label{tab:serverlocation}
\end{table}

We see a general trend of decreasing data rate with increasing (geographical) distance from the server. Clients in Norway record the highest data rates against the measurement server in Norway, followed by Sweden and then Germany (preserving the order of the number of flows established before).
This confirms that server location has a significant effect on measurement results, and having a measurement server as close to the vantage point as possible is of great importance, in order to capture the client-experienced achievable data rate.
\section{Conclusions and Future Work}\label{sec:conclusion}

As data rate builds into a critical selling point in the \ac{MBB} market, there is also a corresponding controversy and an increased interest surrounding it.
There is a plethora of entities which implement measurement tools targeted at end-users, which are essentially alternatives to one another, but cannot be reliably and repeatably compared.

In this study, we focus on \ac{TCP}-based active measurements and provide a configurable tool, MONROE-Nettest within an \ac{EaaS} framework on the \ac{MONROE} platform.
The MONROE-Nettest is designed for compatibility with the largest portion of the available applications and is able to mimic different measurement methodologies.
We provide the first testbed integration of a completely open-source and configurable speed measurement tool, with rich metadata and context information for repeatable large scale measurements in operational \ac{MBB} networks. Such a tool, we believe, is instrumental for the standardization of broadband measurement methods.

Our service can be used to \textit{baseline} measurements which have been conducted with different sets of parameters.
This combines the benefits of crowdsourcing with controlled experimentation, by allowing existing datasets to be jointly used.
To the best of our knowledge, our work is also the first attempt to quantify the effects of different parameters such as number of \ac{TCP} flows, measurement duration, and server location on \ac{DL} data rate measurements in operational \ac{MBB} networks, where detailed results are also provided as open data.

As future work, we plan to develop algorithms that can reduce the data volume consumption while still providing an accurate speed estimate, by adaptively selecting the measurement duration according to network properties.
We also plan to study the time series of data rate jointly with \ac{TCP} metrics and traceroute results, in order to localize bottlenecks in the end-to-end path, and to differentiate whether the observed data rate is limited by the operator's network or within the Internet.

\section{Live Demo}\label{sec:live_demo}
A live demonstration of the MONROE-Nettest will be provided during the workshop, where we will run our tool with different configuration parameters in operational MBB networks and illustrate the results in near real-time.

\bibliographystyle{IEEEtran}
\bibliography{references}

\end{document}